# Untangling Quantum Entanglement

Jeremy L. Fellows[*]

(Dated: October 16, 2005)

**Abstract:** The phenomenon of quantum entanglement is explained in a way which is fully consistent with Einstein's Special Theory of Relativity. A subtle flaw is identified in the logic supporting the view that Bell's Inequality precludes all local hidden-variable theories, and it is shown how EPR-type experiments can be constructed to produce statistical correlation results in a purely classical manner which match exactly the predictions made by quantum theory.

The phenomenon of quantum entanglement ranks among the most perplexing mysteries of modern science, perplexing because the phenomenon – which has been regularly observed and carefully measured in numerous scientific experiments conducted throughout the last several decades – appears to violate one of the most thoroughly validated scientific theories ever developed: Albert Einstein's Special Theory of Relativity.[1] Simply put, quantum entanglement seems to require that action taken at one location have an *instantaneous* effect in a distant location, violating the fundamental principle of relativity theory which asserts that no effect can be instantaneous at a distance -- that nothing in the universe can travel faster than the speed of light.

Throughout his life, Einstein held to the view that quantum theory must be considered an "incomplete" theory – not just because certain aspects of quantum theory seem to conflict with Einstein's own Special Theory of Relativity – but because quantum theory offers no explanation as to *why* Nature works the way that quantum theory predicts it does. Students are simply told that the mathematics of quantum theory allow fantastically accurate predictions to be made of a very wide range of natural phenomena, and that the actual physical processes underlying many of the more bizarre aspects of quantum theory, such as the phenomenon of quantum entanglement, are simply incapable of ever being understood.

Until his death in 1955, Einstein never gave up trying to resolve the apparent conflict between relativity theory and quantum theory, convinced that a more complete theory of quantum mechanics would eventually be developed to explain the phenomenon of quantum entanglement in a way that is consistent with relativity theory. Ten years following Einstein's death, however, a discovery was made which, most physicists now argue, resolves that conflict once and for all. According to the generally accepted modern view of quantum mechanics, that discovery demonstrates that Einstein was simply mistaken -- that quantum entanglement is, in fact, a phenomenon in which action taken at one location has an *instantaneous* effect in another location.

This paper – written on the one hundredth anniversary of Einstein's discovery of the Special Theory of Relativity – exposes an extremely subtle, albeit critical, flaw in the logic underlying the view that Einstein must have been mistaken in his belief that instantaneous action at a distance is impossible. It may be, as Professor Richard Feynman once said, that "no one understands quantum mechanics," [1] but one *can* understand how the logic underlying quantum theory's explanation of the phenomenon of quantum entanglement is fundamentally flawed. Recognizing that flaw leads directly to a theory of quantum entanglement which is fully consistent with relativity theory. To understand why, one must first understand what is so puzzling about the phenomenon of quantum entanglement.

---

[1] Professor Feynman once wrote that "[t]here was a time when the newspapers said that only twelve men understood the theory of relativity. I certainly do not believe there ever was such a time. There might have been a time when only one man did, because he was the only guy who caught on, before he wrote his paper. But after people read the paper a lot of people understood the theory of relativity in some way or other, certainly more than twelve. On the other hand I think I can safely say that nobody understands quantum mechanics." Richard P. Feynman, *The Character of Physical Law* (Cambridge: MIT Press, 1965), p. 129.

1. The Nature and Origin of the Quantum Entanglement Puzzle

The theory of quantum mechanics was developed nearly a century ago in response to a puzzle involving "black body radiation." Classical reasoning suggested that a completely black object (one which would absorb the entire spectrum of electromagnetic radiation) should radiate across the entire spectrum when hot. Experiments with heated black bodies, however, showed that they did not radiate gamma rays, x-rays, or even ultraviolet light. This puzzle was dubbed "the ultraviolet catastrophe."

In 1900, Max Planck solved that puzzle by theorizing that energy must be *quantized.* When electromagnetic radiation is understood as a flow of discreet packages of energy, each of which must be of a size some multiple of a certain minimum value (now referred to as "Planck's Constant"[2]), calculations of the amount of radiation emitted from black bodies exactly match experimental results. For that groundbreaking theoretical insight, Max Planck was awarded the Nobel Prize in physics. His discovery that energy is quantized was spectacularly successful in explaining all manner of natural phenomena that had puzzled physicists for decades, such as why hydrogen atoms absorb and emit electromagnetic radiation only at specified frequencies. As often happens in science, however, the resolution of one puzzle – the mystery of what was causing the "ultraviolet catastrophe" – gave birth to a number of new puzzles. Quantum entanglement is one of those puzzles.

While working on and expanding Planck's quantum theory throughout the early 1900's, physicists came to understand much about the mathematics underlying quantum theory. In 1925, Erwin Schroedinger developed his famous equation[3] which allows scientists to calculate many aspects of the behavior of quantum particles. Because solutions to the Schroedinger Equation are in the form of waves, and because it was clear to Schroedinger that waves could be "added" to one another – both mathematically as well as physically – to produce other solutions to that equation, Schroedinger realized that, in certain special situations, a single solution, or "wave function," could be crafted to calculate the behavior of two or more quantum particles whose attributes are related to, or linked to each other in one way or another.

---

[2]Planck's Constant is equal to 6.6262 x 10$^{-24}$ joule-seconds.

[3]The Schroedinger Equation, shown below, describes the motion of a particle with total energy (E) and mass (m), moving in a single dimension (x), in a region in which there is a potential (V). It is a differential equation, solutions to which take the form of scalar waves like the typical sine wave shown in the drawing below.

$$E\psi = -\frac{\hbar^2}{2m}\frac{d^2\psi}{dx^2} + V\psi$$

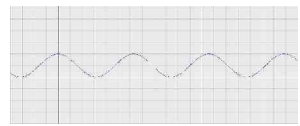

Since that time, scientists have developed a number of different laboratory techniques which allow generation of multiple quantum particles, such as pairs of photons, using processes which insure that the generated pairs have certain qualities that can be described by a *single* Schroedinger Equation. For example, when calcium ions are subjected to radiation, and the energized electrons in the outer orbital shells of the calcium ions are allowed to revert to their normal state, they do so by emitting a *pair* of photons. Well-understood and non-controversial principles of physics dictate that these paired photons must depart the calcium atom in opposite directions and that they must have the same axis of linear polarization.[4] Schroedinger referred to this process of generating paired photons – with related attributes whose motion could be described by a single wave function – as "entanglement."[5]

At about this same time, another physicist, Werner Heisenberg, came to the realization that quantum theory implied a fundamental limitation on the extent to which scientists can measure the position, momentum or certain other attributes of subatomic particles. According to his famous Uncertainty Principle, a discovery for which Heisenberg was awarded the 1932 Nobel Prize in physics, the product of (A) the uncertainty as to the position of a particle, and (B) the uncertainty regarding the momentum of the same particle, must be greater than or equal to Planck's Constant. Based on this principle, as well as certain other discoveries by Heisenberg involving noncommutativity of matrix manipulations,[6] quantum theory arrived at the view that it is not only impossible to precisely *measure* both the position and momentum of a single subatomic particle at the same instant in time, but that subatomic particles *do not **have** a precise position or a precise momentum **until** they are measured*. According to quantum theory, it is the very act of measuring the position (or momentum) of a particle which fixes the particle's position (or its momentum).[7]

---

[4]Light travels through space as an electromagnetic wave. Its undulating electric and magnetic fields are orthogonal to each other, like the two black sine waves shown in each of the two drawings below. The orientation of those fields in *linearly* polarized light is shown on the left. If those fields are offset, as shown in the drawing below on the right, the axis of polarization rotates as the light moves through space. Such light is *circularly* polarized.

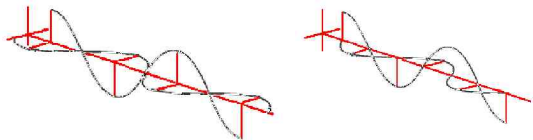

[5]Schroedinger said this about "entanglement": "When two systems, of which we know the states by their respective representation, enter into a temporary physical interaction due to known forces between them and when after a time of mutual influence the systems separate again, then they can no longer be described as before, viz., by endowing each of them with a representative of its own. I would not call that *one* but rather *the* characteristic trait of quantum mechanics." E. Schroedinger, *Proceedings of the Cambridge Philosophical Society*, 31 (1935) 555 (emphasis in original).

[6]Matrix manipulation is an alternate method of calculating the motion of quantum particles which does not use the Schroedinger Equation. When matrices are multiplied together in that process, however, the resulting value depends on the order in which the matrices are multiplied. In other words, when multiplying two matrices together, "A times B" does *not* equal "B times A"! This "noncommutativity" of matrix multiplication lies at the heart of quantum theory's insistence that subatomic particles simply do not *have* precise positions and momentum at the same instant in time, contrary to what classical physics has always assumed. Since a particle cannot have a precise position or momentum until it is measured – at least according to quantum theory – "entangled" particles also do not have precise positions or momentums until they are actually measured.

[7]This is what is sometimes referred to as "collapsing the wave function" to determine a specific value for a particle's location or other attribute from what was previously only a probable estimate of that attribute.

a. The EPR Paper: Quantum Theory vs. the Locality Principle

Einstein was unwilling to accept the foregoing view that subatomic particles *do not have* a precise position or momentum until those values are actually measured.[8] He believed that all subatomic particles *have* a precise position and a precise momentum, even though Heisenberg's Uncertainty Principle may well prevent simultaneous *measurement* of both of those values. To prove his point, and to demonstrate that quantum theory had to be considered "incomplete" as it was understood as of that point in 1935, Einstein enlisted the aid of Boris Podolsky and Nathan Rosen (hereinafter "EPR"), to publish their famous EPR paper,[2] which proposed a clever "*Gedanken Experiment*" – or "thought experiment" – involving entangled subatomic particles.

In that paper, Einstein and his associates reasoned as follows: If a pair of entangled subatomic particles were generated, and the position of one of those particles were precisely measured, a precision measurement could then be made of the *momentum* of the *second* particle. That would allow one to then calculate mathematically the momentum of the *first* particle (as well as the *location* of the *second* particle) without having to actually measure those values, thereby cleverly circumventing quantum theory's assertion that it is physically impossible to know both the location and the momentum of a single particle at the same moment in time.

The EPR paper constituted a full frontal attack on the very foundations of quantum theory. In response to that attack, Niels Bohr – one of the greatest proponents of, and contributors to, quantum theory – pointed out that the so-called "EPR paradox" was entirely predicated on the aforementioned fundamental principle of relativity theory which states that action taken at one location cannot have an instantaneous effect at some other location, a principle often referred to as the *Locality Principle*. Bohr struck back at the EPR paper by arguing that the Locality Principle simply must not be valid. In other words, according to Bohr, measuring the location of one of a pair of entangled photons *does* have an instantaneous effect on the other entangled photon, even though it may be located a great distance away. Bohr dismissed the EPR paradox by saying that the Locality Principle simply must not be part of our reality, despite Einstein's belief that it should be.

Until his death in 1955, Einstein held to his view that no reasonable definition of reality can allow the physical attribute of one subatomic particle to depend on the process of measuring the physical attribute of some other "entangled" subatomic particle in a distant location. Any such "spukhafte Fernwirkungen" or "spooky action at a distance" – as Einstein called it – cannot be part of our physical reality. This seemingly irreconcilable stalemate between Einstein's firm belief in the Locality Principle and nearly everyone else's opposing faith in quantum theory lasted for decades until John Bell developed a mathematical theorem in 1964 designed to resolve that stalemate once and for all -- to determine whether the Locality Principle is real. That theorem -- "Bell's Inequality" – is a discovery which has been referred to as "the most profound discovery of science."[9]

---

[8] When discussing this aspect of quantum theory, Einstein once commented, "I think that a particle must have a separate reality independent of the measurements. That is, an electron has spin, location and so forth even when it is not being measured. I like to think that the moon is there even if I am not looking at it."

[9] *Process Studies*, pp. 173-182, Vol. 7, Number 3, Fall, 1977.

b. Bell's Inequality - The Test to Determine if the Locality Principle is Real

In his 1964 paper entitled "On the Einstein Podolsky Rosen Paradox,"[3] Bell interpreted Einstein's EPR paper as advancing the argument that the mathematics of quantum mechanics must be supplemented by additional variables in order to make quantum theory consistent with the Locality Principle.[10] Bell then endeavored to show that if the statistical results of a "correlation experiment" (discussed in detail below) could be shown to violate his inequality theorem – as quantum theory predicts will happen – the Locality Principle must be invalid. Ironically, Bell appears to have believed that Einstein's intuition was correct, and that the statistical results of "correlation experiments" – when eventually conducted and evaluated using his inequality theorem – would *not* violate that theorem, thereby supporting Einstein's view that quantum theory is an "incomplete" theory. As noted above, however, the statistical results of those experiments, when they were eventually performed, were not at all what Bell expected. On the contrary, those results have been interpreted as proving rather conclusively that Einstein must have been wrong and that the Locality Principle must be *invalid*. To fully understand why those experimental results, as surprising as they are, do *not* justify the conclusion that Einstein was wrong, it is necessary to understand the extraordinarily simple logic which underlies Bell's somewhat complex theorem.

Since it was first published in 1964, Bell's Inequality has been expressed mathematically in a number of very different ways,[11] but the basic logic underlying of each of those mathematical expressions is exactly the same. Most commonly, Bell's Inequality is written as some version of the following equation:

$$n[X,-Y] + n[Y,-Z] \geq n[X,-Z]$$

That equation, however written, expresses a relationship between three related quantities (X, Y and Z). Stated most simply, Bell's Inequality says that -- for any three categories or groups of any kind of items or objects of any sort one wishes to consider -- **the number which will fall into the first category, but not into the second category, plus the number which fall into the second, but not the third category, will always be equal to or greater than the number which fall into the first, but not the third category**.

This fundamental mathematical relationship between categories remains valid, according to Bell's theorem, regardless of what you choose as categories, and regardless of the number of items which fall into any of the three categories. For example, if Category X is the number of men in the room, Category Y is the number of people in the room who are Egyptian,

---

[10] In his EPR paper, Einstein actually did not speak in terms of "hidden variables." He simply expressed the view that quantum theory was "incomplete" because it could not account for the results of the his "thought experiment." It was John von Neuman, a well-respected mathematician and contemporary of Einstein's, who had written a paper suggesting that additional "hidden variables" might resolve the EPR paradox. Bell was quite familiar with von Neuman's paper, having proven in a paper finally published in 1966 that von Neuman's analysis in that regard was flawed.

[11] According to Günther Schachner in his December 12, 2003, article entitled "The Structure of Bell Inequalities," any Bell Inequality can be written in the following form:

$$\left| \sum_{k=0}^{2^n - 1} b_k E(k) \right| \leq 2^n$$

and Category Z is the number of people in the room who had turkey sandwiches for lunch today, Bell's Inequality tells us that the number of men in the room who are not Egyptian, plus the number of Egyptians in the room who didn't eat turkey sandwiches for lunch today, will be equal to or larger than the number of men in the room who didn't eat turkey sandwiches for lunch today. While this example of Bell's Inequality in action may seem pointless and unimportant, the formula has proven to be of immense importance in determining whether there is support for Einstein's belief that the Locality Principle is a valid principle, or whether the Locality Principle is at odds with quantum theory, and therefore, invalid.

The easiest way to understand the logic underlying Bell's Inequality is to consider how one might best portray the relative size and overlapping relationships of three different kinds or categories of things. The simplest and most readily understandable approach is to show that information graphically using Venn diagrams like the one shown below.

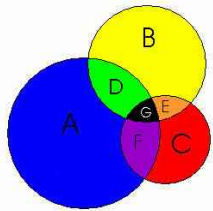

Using this method of identifying categories and subcategories, we can now refer to the first *main* category – i.e., the largest circle – as the sum of the blue subcategory, plus the green subcategory, plus the purple subcategory, plus the black subcategory in the center.

In short, all of Category X (the large circle) can be represented by the equation: "A+D+F+G". Similarly, Category Y (the second largest circle) can be represented by the equation: "B+D+E+G". Finally, Category Z can be represented by the equation: "C+E+F+G".

Following this colored subcategory approach, the quantity on the *right* side of Bell's Inequality (i.e., the part following the *greater than or equal to* sign) can be portrayed graphically using this method as shown below.

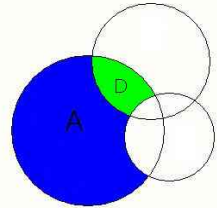

This two-colored area represents items which are members of Category X, but *not* Category Z, and can be designated by the expression "A+D". The *first* half of Bell's Inequality (i.e., the part on the *left* side of the equation, before the *greater than or equal to* sign) encompasses the quantity of items which fall into Category X, but not into Category Y, *plus* the number which fall into Category Y, but not into Category Z. Coloring in the same Venn diagram using the foregoing color scheme, we see that the left side of Bell's Inequality can be represented graphically as follows:

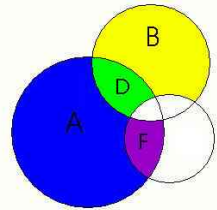

The left-hand side of equation can therefore be designated as "(A+F) + (B+D)". Those letters can be rearranged as "A+B+D+F" or even as "(B+F) + (A+D)". Putting both sides of Bell's Inequality together, we have the following statement in terms of letters representing areas in our Venn diagram:

$$(B+F) + (A+D) \geq (A+D)$$

The simple logic behind Bell's Inequality can easily be understood by noting that the letters A and D appear on *both* sides of the foregoing equation. If we subtract A and D from both sides of the equation, which we are allowed to do here as in any equation, we are left with an equation stating that B plus F is always greater than or equal to zero. If B or F have any size at all, this means something is always greater than or equal to nothing. How profound!

When one looks at the logic behind Bell's Inequality in this way, it is easy to see why it makes no difference whether the size of any of the three subject categories is greater than the size of either of the other two categories. The inherent logic underlying Bell's Inequality is equally valid, no matter how large any of the categories are, and no matter how many items in any particular main category are also members of one or more of the other two main categories. The logic which underlies Bell's Inequality (i.e., that something is always greater than or equal to nothing) remains valid regardless of any change one might wish to make to the relative size of any or all of the three main categories, or the extent to which they overlap one another.

For example, in the diagram shown below, the size and relative relationship of the three main categories have been significantly altered, yet Bell's Inequality remains equally valid. B plus F is still greater than or equal to zero!

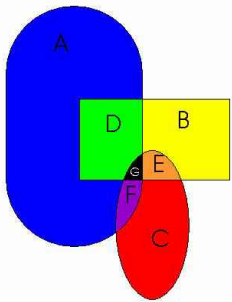

It makes no difference whether one shrinks the size of one or more of the three main categories to the extent that they are eliminated entirely, or arranges things so that one of the main categories is completely contained within another category. Bell's Inequality remains equally valid. For example, in each of the diagrams shown below, there are *no* objects which fall into Category Z (the small ellipse) which do not also fall into Category X (the large pill-shaped area).

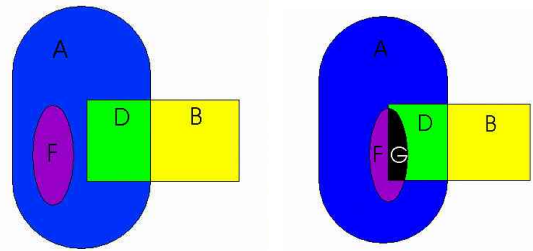

Nevertheless, Bell's Inequality remains valid for the foregoing diagrams, since there is simply no way to *ever* draw a figure this way in which the blue, purple, yellow and green areas – when added together – represent less area than the sum of just the blue and green areas. Viewed this way, it is easy to see that Bell's Inequality is simply tautological: It can *never* be violated, unless it is used in a way that violates one of its implicit limitations.[12] The fact that Bell's Inequality is tautological is why it's so shocking to learn that the results of statistical correlation experiments (discussed in detail below) conducted in the laboratory in the last few decades in an attempt to resolve the aforementioned dispute between Einstein's belief in the Locality Principle and the conflicting assertions made by quantum theory – appear to *violate* Bell's Inequality.

---

[12] One such limitation is that there be no change in the size of, or relationship between, any of the subcategories merely as a result of measuring the size of a category, such as would happen if the categories were (1) the number of magicians in the room; (2) the number of people in the room with a hat on their head; and (3) the number of people in the room with a rabbit under their hat. Removing hats to look for rabbits *alters* the size of category number (2).

To understanding why the results of those experiments should *not* surprise us, or cause us to believe that entangled particles can affect each other *instantaneously* as quantum theory asserts, it is first necessary to understand how Bell's theorem has been wielded in an attempt to prove that Einstein must have been mistaken – to prove in other words that Nature is "non-local."

c. The Improved Bell's Inequality and Its Use in Correlation Experiments

The EPR paradox asserted that quantum theory must be considered incomplete because the results of Einstein's "thought experiment" are in direct conflict with quantum theory. Unfortunately, at the time the EPR paper was written, limitations inherent in the laboratory equipment and experimental techniques then available made it impossible to actually perform such an experiment. Some decades later, however, in 1969, John Clauser, Mike Horne, Abner Shimony and Richard Holt made such experiments possible by improving Bell's theorem, rewriting it to eliminate a certain limiting assumption implicit in the original formula. As rewritten, the so-called "CHSH" version of Bell's Inequality[4] looks like this:

$$B(a_1, b_1, c_2, d_2) = |q(b_1,d_2) - q(a_1,d_2)| + |q(b_1,c_2)+q(a_1,c_2)| < 2$$

This improvement in Bell's original formula, and certain advances in laboratory and experimental techniques, finally made it possible by the 1980's to conduct actual laboratory experiments designed to determine whether Bell's Inequality is, in fact, violated by the results of experiments using entangled photons.[13] Those experiments are usually referred to as "correlation experiments" since they are designed to study the extent to which a measurement of the polarization of a photon, is correlated with the measurement of its entangled counterpart.

The results of those experiments, as currently interpreted, are cited in support of the nearly unanimous view that Einstein must have been wrong -- that the universe is, in fact, *non-local*, meaning that measurement of a photon in one location does have an *instantaneous* effect on its entangled counterpart, regardless of where the other photon may be located when the measurement is made.

As will be demonstrated below, however, the accepted, orthodox interpretation of the results of those correlation experiments is based on a fundamental misunderstanding as to what causes the phenomenon of "entanglement" observed in those experiments. To fully understand how the results of those experiments have been misinterpreted, and why Einstein was *not* wrong, one must first understand how those experiments are actually performed, and what they attempt to show.

---

[13]Einstein's thought experiment involved measurements of the location and momentum of entangled massive particles – rather than measurements of the polarization of entangled photons – but the relative ease of measuring photon polarization, as contrasted with the relative difficulties inherent in measuring a particle's exact location or momentum, has caused most researchers to use entangled photons when conducting such experiments. No one has suggested, nor is there any reason to believe, that the bizarre outcome of those experiments would be any different if the experiments were conducted using electrons, positrons, or any particles other than photons.

-8-

i. Correlation experiments - How they work and what they measure

Correlation experiments designed to use entangled photons create pairs of photons in a number of different ways. As mentioned above, one common method – called the atomic cascade method – creates pairs of entangled photons by subjecting calcium ions to radiation, and then allowing the energized electrons in the outer orbital shells of the ions to revert to their normal state. When they do so, they simultaneously emit pairs of photons which, according to well-accepted laws of physics, *must* depart the atom in opposite directions with precisely the same axis of linear polarization. Although there are other methods of generating entangled photons, the correlation experiments which employ those other methods produce the same overall statistical results. In other words, the particular method of creating entangled photons has no observable effect on the statistical outcome of correlation experiments.

The experiments are typically constructed so that the photon source emits entangled photons in opposite directions, which then pass through separate polarizers designed to act as filters, allowing passage of only those photons with a specified polarity. The polarity of emitted photons is random in general, but polarity of entangled photon pairs is *correlated*. In other words, one cannot predict what polarity any particular photon will have – considered independently from all other photons in the experiment, but once the polarization of any particular photon is measured, well-accepted principles of physics guarantee that the measured photon's entangled pair will have exactly the same axis of polarization.

The polarizing filters can be rotated to any fixed angle to permit passage of photons with any specified polarity. After passing through their respective polarizers, photons which are not blocked[14] by the polarizers are detected by separate photon detectors. The photon detectors are connected in turn to a device designed to monitor whether two photons are detected *simultaneously* at both detectors. If two photons are detected simultaneously (i.e., one or both are not blocked by the polarizer), they are counted as a "correlated" pair for the specified angles. A simple experimental setup designed to measure the correlation of photons passing through polarizing filters set to a 315-degree angle on the left side of the apparatus and a 119-degree angle on the right side would look something like the diagram below.

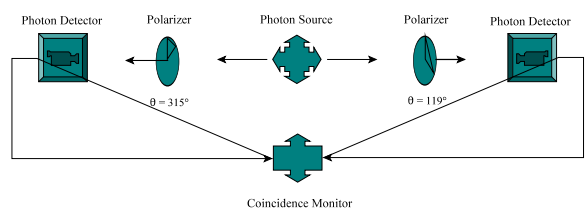

Once the polarizers are set to the desired angles, the experiment is allowed to run until a very large number of photon pairs are counted and their correlations are calculated for the specified angles. The resulting statistics can then be analyzed to determine whether the results of the experiment match what is predicted. Correlation experiments of this nature are quite fascinating, because quantum theory predicts results which differ from the results we are told to expect if Einstein were correct that the Locality Principle governs the outcome of such experiments.

---

[14]More elaborate experimental setups are possible in which the polarizing filters are replaced with birefringent crystals which separate incident light into oppositely polarized rays of light according to the axis of polarization of the incident light. Such setups allow monitoring of *two* channels, permitting calculation of correlation statistics for both those photons which would normally pass through a polarizing filter, as well as those which would otherwise be blocked by the filter. The overall statistical results produced by these more elaborate correlation experiments are essentially the same, albeit somewhat more precise.

ii. Statistical Results as Predicted by Classical Physics vs. Quantum Theory

Because the polarizations of entangled photon pairs are identical, the results of such experiments predictably show 100% correlation whenever the polarizers are set to the same angle, regardless of which theory applies (i.e., quantum theory or the theory which says that the universe is *local*, in that nothing can travel faster than the speed of light). Similarly, if the two polarizing filters are set to angles *perpendicular* to one another, the results are predictably completely *anti*-correlated, which is equally consistent with both theories. However, as the so-called "theta angle" between the two polarizing filters' respective orientation changes from zero degrees to 90 degrees, quantum theory predicts results which differ from the results predicted by the accepted view of classical physics. The accepted view of classical physics predicts a straight-line or *linear* graph of the outcome of this type of experiment, similar to the one shown below. Why this is so is discussed in more detail below.

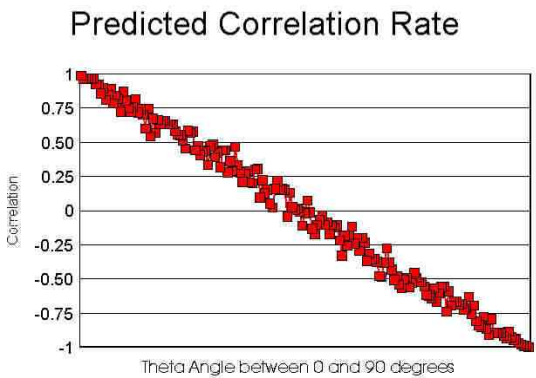

Quantum mechanics ("QM"), on the other hand, makes the prediction that a graph of the overall outcome of a correlation experiment using spin-1 particles like photons should produce a *nonlinear* sinusoidal curve similar to the example depicted below.

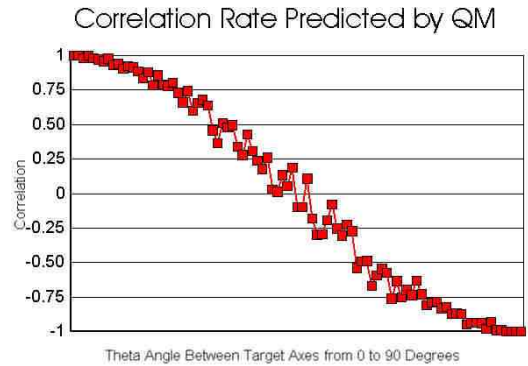

To date, quantum theory has not offered any explanation as to *why* Nature behaves this way. The mathematics of quantum theory simply predicts that correlation experiments of this nature will produce the *nonlinear* curve shown in the graph depicted above.

Since Bell's Inequality allows one to analyze the correlations among *three* categories, the typical correlation experiment is generally configured to calculate the rate of correlation between entangled photons at three separate angles – including the two angles where the differences between the predictions of the opposing theories are the greatest. Those differences are greatest where the axes of polarization of the filters are oriented with respect to each other at a so-called "theta angle" of 22.5 degrees, and 67.5 degrees. The third optional polarizer setting is typically zero degrees, where the polarizing filters are set parallel to each other. A count of the number of correlated photons during a set period of time is then made three separate times: the first with the filters set parallel to one another (the zero degree theta angle), then with the filters set 22.5 degrees apart, and finally with the filters set 67.5 degrees apart. The data showing the relative number of correlated photons at those three settings is then examined to determine whether the results violate Bell's Inequality.[15]

The same basic method would theoretically be used for correlation experiments using "spin-½" particles like electrons or positrons, although the experiment would be configured to detect whether the particles were "spin-up" or "spin-down", rather than what axis of polarization the particles might have. Instead of using polarizing filters, the experimental apparatus would make use of Stern-Gerlach magnets to detect particle spin. Correlation experiments using spin-½ particles like electrons or positrons would be expected to produce results which differ in one respect from experiments using entangled photons, in that the spin of such spin-½ particles would theoretically be correlated whenever the targets are positioned *opposite* one another (i.e., at a theta angle of 180 degrees from one another), and *anti*-correlated when the theta angle is set to zero degrees (i.e., when the targets have the same orientation).

This is because accepted laws of physics require entangled spin-½ particles to have spins which are *opposite* from one another. Because of this difference, the accepted view of classical physics would predict a linear graph of the statistical outcome which slants *upwards* as the theta angle between the polarizing filter targets is increased, as shown in the following example:

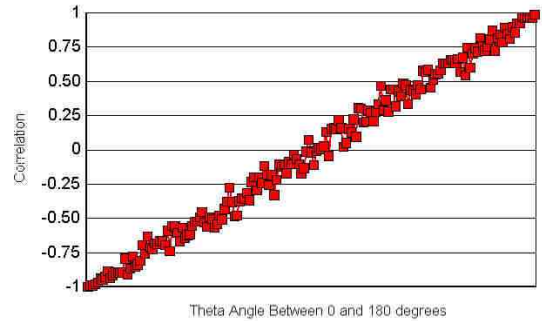

Quantum theory, on the other hand, still predicts a *nonlinear* graph of the statistical results of such an experiment, with the correlation data aligned along a sinusoidal curve similarly flipped on its horizontal axis. Regardless of which type of correlation experiment is conducted, however, one thing is clear: If the correlation experiment is set up as usual to measure **three** optional target orientation settings, there are only **eight** ways in which pairs of entangled particles can be generated so that the spins of the electron/positron pairs – or the polarization of pairs of photons – will be properly correlated when the two targets happen to be set to the same optional setting.

For example, there are only eight ways in which pairs of entangled spin-½ particles can be created so that the pairs will always have opposite spin, regardless of how the target settings are configured. The table on the next page shows all of the eight configurations in which a pair of entangled spin-½ particles can theoretically be "preprogrammed" for three specified target angles so that the left particle spin will always be *opposite* of the right particle spin for the same specified angle.

---

[15]The CHSH version of Bell's Inequality says that if the statistical results, when fed into that formula, produce a number greater than two (2), the results are consistent with the predictions of quantum mechanics and inconsistent with any theory relying on the Locality Principle. Using the specific angles chosen in the foregoing example (i.e., zero degrees, 22.5 degrees and 67.5 degrees), quantum mechanics predicts that the statistical results should *violate* Bell's Inequality, producing a value equal to two times the square root of two (i.e., $2 \times \sqrt{2}$ or approximately 2.82).

|   | Left Particle Spin | | | Right Particle Spin | | |
|---|---|---|---|---|---|---|
|   | 0° | 22.5° | 67.5° | 0° | 22.5° | 67.5° |
| 1 | Up | Up | Up | Down | Down | Down |
| 2 | Up | Up | Down | Down | Down | Up |
| 3 | Up | Down | Up | Down | Up | Down |
| 4 | Down | Up | Up | Up | Down | Down |
| 5 | Up | Down | Down | Down | Up | Up |
| 6 | Down | Down | Up | Up | Up | Down |
| 7 | Down | Up | Down | Up | Down | Up |
| 8 | Down | Down | Down | Up | Up | Up |

Since entangled pairs of particles in correlation experiments are presumably generated in a completely random manner which can't favor any of the foregoing eight optional configurations, classical theory has always assumed that each of the foregoing eight configurations should occur with the same overall frequency. This assumption has led to calculation of the following predicted statistical probabilities of finding correlated spins at the specified target settings:

| Probability of Correlated Spins | | Right Target Angle | | |
|---|---|---|---|---|
| | | 0° | 22.5° | 67.5° |
| Left Target Angle | 0° | 0 | ½ | ½ |
| | 22.5° | ½ | 0 | ½ |
| | 67.5° | ½ | ½ | 0 |

These statistics show that there is zero chance of correlated spin (i.e., the spin results are perfectly *anti*-correlated) whenever the targets are set to the same angle. Otherwise, there is a 50% chance (i.e., a probability of ½) that both particles will have the same spin for every other possible way in which the two targets can be positioned at any of the three specified angles.[16] Adding together all of the foregoing probabilities for each of these nine possible ways in which the targets can be aligned, it is easy to show that there should be an *overall probability of only **one-third (⅓)*** that the particles will be determined to have the same spin.[17] Quantum theory, on the other hand, predicts a much greater likelihood of correlated spin. Quantum theory predicts that there will be an *overall probability of **one-half (½)*** that the particles will be determined to have the same spin.

As mentioned above, when it finally became possible to actually perform photon-based correlation experiments in the laboratory using photons, the experiments nearly always produced results just as predicted by quantum mechanics![5] Those results are deeply perplexing, however, because the results seem to suggest a greater degree of interdependence between spatially separate events than can be accounted for by classical causality theory -- at least as presently understood.

To understand why orthodox classical causality theory can't account for those results, and why any other explanation suggests that the inanimate particles used in those experiments appear to be *communicating with each other* – passing information to each other faster than the speed of light – one must look at how those results have been analyzed and explained.

---

[16] For example, if the left target is at 22.5° and the right target is at 67.5°, spin will be matched in four out of eight optional configurations. The same can be said where the left target is at 0° and the right target is at 22.5°. Conversely, where both targets are positioned to the same optional angle, particle spin never matches.

[17] Of the nine optional configurations, there are six which each have a predicted probability of one-half, according to the accepted view of classical physics. Six times one-half totals to three, and three ninths is equal to an *overall* probability of statistical correlation of one third.

iii. How the Bizarre Results of Correlation Experiments Have Been Explained

The belief that entangled photons in the typical correlation experiment must be passing information to each other faster than the speed of light is based on the following analysis: If there are three optional target angles (i.e., 0°, 22.5° and 67.5°) the only way randomly generated particles can generate correlation statistics which *exceed* the overall ⅓ probability calculated in the chart shown above on page 12 is if the photons somehow communicate with each other to insure that their polarization is matched more frequently than would otherwise occur if they were unable to communicate with each other and their respective polarizations were determined completely randomly.

According to the foregoing analysis, increasing the statistical probability of correlated results from an overall likelihood of one-third (⅓) to an overall likelihood of one-half (½), requires that at least one of each pair of entangled photons somehow be "aware" of how many optional target settings there are (i.e., three in the example discussed above), and what polarizations (i.e., nearer to the horizontal than the vertical, etc.) will guarantee that the polarizations of the photons will be appropriately correlated (or anti-correlated), regardless of how the targets happen to be positioned when the photons reach their targets.

If there are three optional target settings, at least one of each pair of photons must therefore be armed with no less than three bits of information: (1) How it must orient itself if the other target is set to optional position A; (2) How it must orient itself if that target is set to optional position B; and (3) How it must orient itself if that target is set to optional position C.

What is more important, before the photon armed with that knowledge reaches its own target, *it must learn how the other target is positioned*.[18] Only with that information can the photon know how to adjust its own polarization in order to properly correlate its orientation with that of its paired photon so as to generate the correlation statistics required to meet the predictions of quantum mechanics. Without information about the orientation of the other target, the statistical results would theoretically be based on pure randomness, which would presumably produce the *linear* correlation statistics mentioned above (i.e., those with an overall probability of correlation of only one third).

The central problem is this: How can information concerning the orientation of one target get communicated to the photon headed toward the *other* target? In a correlation experiment using photons of light, there is no way for information about target positioning on the left side of the apparatus to be communicated to a photon moving toward the target on the right side of the apparatus. Even if there were such a process, as yet unknown to science, that information could not travel fast enough to reach the particle on the right before that particle arrives at its target -- at least not without violating Einstein's Special Theory of

---

[18] Each photon is presumably "aware" of its paired photon's orientation, since the orientation is the same in the case of photons, or the opposite (i.e., spin-up or spin-down) in the case of spin-½ particles like electrons or positrons. Neither particle can be similarly "aware" of the orientation of either target, especially the one on the opposite side of the apparatus. A particle becomes "aware" of the orientation of its *own* target only when it reaches it.

Relativity, which limits speeds to the speed of photons themselves.[19]

Quantum theory sidesteps that problem by simply asserting that entangled particles don't *have* a particular spin or polarization until the spin/polarization of one of the entangled particles is actually measured. According to quantum theory the act of measuring fixes the spin/polarization for *both* entangled particles at the same instant in time. Einstein believed, however, that this explanation still necessarily means that action taken at one point in space must have an instantaneous effect on a particle in a distant location, and that such an effect is prohibited by the Special Theory of Relativity.

When trying to resolve this conundrum, researchers have reasoned that if it is impossible to transmit the required information through space instantaneously, because of the speed limitation rule of the Special Theory of Relativity, entangled particles must somehow be *created* with all of the information they need to know *in advance* – telling them how to respond to every potential target position in order to produce the inexplicably high correlation statistics measured in such experiments. This theory – that information about target positioning is somehow hidden within the entangled particles themselves from the moment they are created – is often referred to as the "hidden variable" theory.

Bell developed his inequality theorem in order to determine whether any such "hidden variable" theory could ever account for the inexplicably high correlation statistics predicted by quantum theory. One of the tacit assumptions he made when developing that theorem, however, is that *the hidden variables producing the bizarre statistical results – if they exist – must be intrinsic to the entangled particles themselves*. Put another way, Bell simply assumed that the experimental apparatus used to measure the entangled particles plays a completely passive role, having no significant effect on the resulting statistics. This tacit Passive Apparatus Assumption, in turn, leads directly to the additional implicit assumption that, in order for an entangled photon to "know" whether its axis of polarization should be at one angle or another when it reaches its polarizing filter, it must "know" – for each of the different optional polarizer angle settings – how it must respond when it arrives at any one of those optional settings. In other words, if the correlation experiment allows for three optional polarizer settings, the minimum required bits of hidden information must total no less than three.

As will be shown below, there is no logical basis for the first of these two critical assumptions, and the second – the Minimum Information Assumption – is demonstrably false. These erroneous assumptions are, in fact, the source of the mystery surrounding quantum entanglement! The failure to recognize the falsity of those two assumptions is precisely what has misled the entire community of physicists and researchers exploring this extraordinarily important area of science into believing that the statistical results of correlation experiments are necessarily inconsistent with Einstein's Special Theory of Relativity and the Locality Principle.

---

[19] In Bell's paper, he recommended that correlation experiments be configured so that the target positions are not chosen until *after* the paired particles have already begun moving towards their target. Extremely rapid switching experiments which have been done in ways that appear to satisfy that concern still produce results consistent with the predictions of quantum mechanics. Supporters of quantum theory have taken the position, therefore, that those experiments prove that Einstein must have been mistaken -- that a measurement of a photon's polarization performed at one location *does* have an *instantaneous* effect in some distant location.

2. The Flaw in the Minimum Information Assumption

But how, one might ask, can a photon know how to respond to one of three separate optional polarizer settings armed with *less than* three bits of information? The answer to that question – which is revealed in the following illustration – is the key to understanding the puzzle of quantum entanglement.

Two brothers, who live in different cities, decide to demonstrate the point being made here to their friend. The brothers ask the friend to keep statistics showing what each brother has for lunch each day for an entire year. The brothers agree that they will have lunch each day in one of any of the several restaurants which are part of one of the following three different restaurant chains in their respective cities: Dennys®, Elmers®, or IHOP®. The particular restaurant they will each be required to eat in will be chosen at random by the friend under circumstances absolutely insuring that neither brother can know which restaurant his brother will be eating in on any particular day before the meal takes place. Just *before* the selection of restaurants is made each day and the brothers depart to have lunch, the brothers are permitted to speak briefly with each other.

Subsequent examination of the friend's compiled statistics show that each brother's choice of what he ate for lunch on a daily basis, as well as where he ate, appears entirely random. One day they ate one thing, the next day something else. There is no discernable pattern at all to where they ate or what they ordered. Nevertheless, when comparing what one brother chose to eat for lunch with the choice made by the other brother, the record shows that the brothers *always* chose the same thing to eat *whenever they happened to wind up eating at the same restaurant on the same day.*

In other words, if they both happened to eat at a Dennys® restaurant on a particular day, they invariably ordered precisely the same thing for lunch. If, on the other hand, they ate at different restaurants on a particular day, their choices were not similarly correlated, but instead completely random. The friend assumes that this result is possible only if the brothers told one another three things before leaving for lunch each day: (1) What they would be having for lunch if they wound up at a Dennys®; (2) What they would be having for lunch if they wound up at an Elmers®; and (3) What they would be having for lunch if they wound up at an IHOP®. Then, after leaving for lunch, they had to somehow communicate with each other to tell each other where, in fact, they had been sent to eat that day. Otherwise, the friend reasons, the brothers could not possibly have insured that they would eat the same thing whenever they wound up eating in identical restaurants on the same day. The friend was wrong.

[Before reading further, the reader may wish to pause to try to identify the subtle logical flaw in the friend's analysis. Such an exercise can help show why it has always been assumed -- albeit quite mistakenly -- that his analysis is logically sound when using Bell's Inequality to analyze the results of correlation experiments like those discussed above.]

Neither brother knew what the other brother would be ordering for lunch. Neither brother knew which of the three optional restaurants the other brother would be eating lunch in until long after they had finished eating their meals and returned home. They didn't need that information in order to guarantee that they would eat the same thing if they happened to wind up eating in the same restaurant. They had another, simpler, method: they ordered from the *menu*, and before leaving to go to lunch each day they each picked a number at random and shared that number with their brother. Armed only with the two randomly chosen daily numbers, the brothers were able to insure they would eat exactly the same thing if they happened to eat in the same restaurant that day. They accomplished that by using the two numbers – together with the menu – to select their meal.

For example, if the random number chosen by the one brother was 6, and the random number chosen by the other brother was 4, they would both measure 6 inches down and 4 inches across on their respective menus and order whatever they found there. This simple method guaranteed that the statistics would turn out as they did, since the menus at all of the Dennys® restaurants in those cities were exactly the same, albeit completely different from the menus in any of the other restaurants. Similarly, the menus at all of the Elmers® restaurants in those cities were exactly the same, albeit different from the menus used in the IHOP® or Dennys® restaurants. The brothers didn't need to know in advance what each other would be ordering for lunch as long as they had some mechanism to insure that they would necessarily order the same thing if they wound up eating in the same restaurant. The menus – and a daily pair of shared random numbers – provided that mechanism.

But what relevance does the use of a menu and shared random numbers in the foregoing illustration have to do with correlation experiments involving photons and polarizers? Well, it appears likely that Nature employs a somewhat similar mechanism when entangled photons are subjected to measurement in the typical correlation experiment. Instead of using shared random numbers, however, as in the foregoing illustration, entangled photons simply "use" their matched orientation in three-dimensional space perpendicular to their direction of travel as the key to determining whether they will pass through their respective polarizing filters. The menu in the foregoing illustration is analogous to the polarizing filters themselves. To understand this concept more clearly, and to see why Bell should not have tacitly assumed, as he did, that the experimental apparatus cannot play more than a purely passive role in producing the bizarre statistical results in correlation experiments, consider the following illustration involving a knife-throwing demonstration, which is set up to function as much as possible like a macro-sized correlation experiment.

a. The Knife-Throwing Demonstration – How the Apparatus Can Effect Results

Imagine a circus act featuring two knife-throwing twins, who perform while strapped to opposite sides of a circular disk approximately six-foot in diameter – and which can be rotated in place – positioned in the middle of a typical auditorium stage. The twins are strapped to the disk head-to-toe *vis a vis* each other, each facing an identical flat target an equal distance away at stage-right and stage-left, respectively. The targets, which are themselves mounted on individual disks which also can be rotated, each have an identical narrow vertical slit – an opening barely wider than the width of the knives (which are incredibly thin) and just as long as the length of the knives – positioned so that each twin sees a target which looks identical from the viewpoint of each twin. The narrow slit in the target looks like the example shown at the top of the following page.[20]

---

[20] The drawing shows a target from the view point of the twin facing that target. The faint web pattern is merely the background of the circular target. The opening is the angled slit through the center of the web. The length of that slit is equal to the length of the knives, which are all identical to each other.

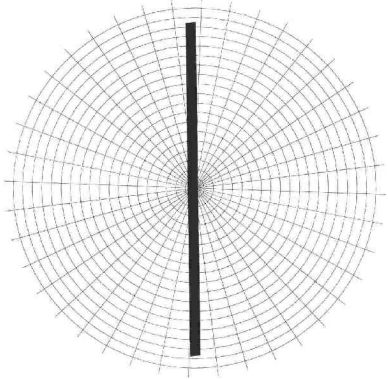

Part One of the act begins with a preliminary demonstration of knife-throwing skill in which the twins each throw a knife simultaneously through the narrow slits in the targets. The knives are identical to each other in all respects. The large circular disk to which the twins are affixed, as well as the disks on which both targets are mounted, are each rotated to a variety of different positions before each successive pair of knives are thrown.

The twins consistently succeed in throwing all of their knives through the narrow slits, regardless of their positions and the relative positions of their targets, but that feat is easily attributed to the twins' great knife-throwing skill, which allows them to adjust the angle of roll[21] of their knives to precisely match the particular angle at which the narrow slit in their own target happens to be positioned at the time their knife is thrown, insuring that their knives always pass through their respective openings.

Then, in Part Two of the act, the twins are both *blindfolded* and two small curtains are positioned between the audience and each of the targets to block the audiences' view of the surface of the targets. The targets with the narrow slits are then replaced with new targets with openings which cannot be seen by the audience. The center disk and both target disks are then set rapidly spinning.

*Despite their blindfolds*, the twins are still consistently able to throw all of their knives through the openings in their respective targets. The secret to this trick is revealed when the small curtains blocking the audience's view of the targets are pulled back: the openings in the new targets are large *circular* openings, each with a diameter equal to the length of the knives. Regardless of what roll and pitch the knives have in flight, they all easily make it through the openings as long as they are thrown straight at the center of the targets, which the twins are very skilled at doing, even when blindfolded. The twins' view of the targets with the large circular openings looks similar to the drawing shown below.

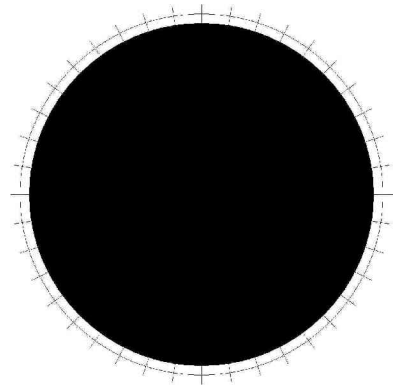

Part Three of their act consists of an identical knife-throwing demonstration, but using targets with circular openings which are only *half* as wide as the length of the knives. Those targets look like the drawing shown at the top of the following page.

---

[21] The term "roll" here means the same thing as an airplane's "roll." For an airplane, it's the degree to which the airplane's wingtips are tipped clockwise or counter-clockwise in the planar surface perpendicular to the line of travel, causing the airplane to bank to the left or right. The term "pitch" (used hereinafter) on the other hand, is the degree to which the nose of an airplane is tipped up or down, causing the airplane to dive or climb.

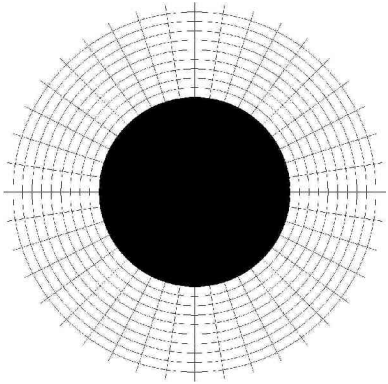

With this change, the twins -- who remain blindfolded -- are no longer able to throw *every* pair of knives through the smaller circular openings in their respective targets. The knives which manage to pass through these smaller openings are the ones thrown with an angle of pitch sufficiently perpendicular to the planar surface of the target (like a dart thrown at a dartboard) to permit the knives to pass through the hole without striking the edge of the target opening and being deflected away. Although the pitch of any *pair* of thrown knives changes randomly from one pair to the next, the twins throw each pair of knives with exactly the same angle of pitch.[22]

Because each of the two knives in any pair of knives are thrown with identical pitch, the twins' success in throwing knives through the smaller circular target openings is precisely correlated: If one twin fails to throw his knife through his target opening because the pitch of his knife is too perpendicular to its line of travel in flight, his twin also fails -- for the same reason. Because the openings in the targets used in this part of the act are exactly half as wide in diameter as the overall length of the knives, the blindfolded twins fail to throw their knives through the openings exactly half of the time. Their success in throwing knives through these smaller circular openings is otherwise completely random in all respects.

In Part Four of their act, the twins replace the small circular targets with new targets which have the same circular opening, but which have colored backgrounds as shown in the drawing below.

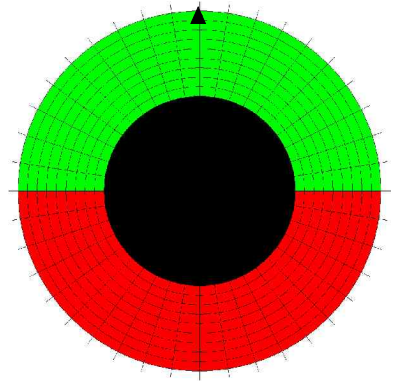

The twins' assistant then places a small triangular black mark at the center of the top edge of both target openings, just like small mark shown at the top of the drawing above. The twins announce that their assistant will keep detailed statistics showing how often they are able to throw their knives so that the *tip* of each knife is closer to the green side than the red side at the moment the tip of the knife reaches the opening. Knives which pass through the opening with their tip nearer the green side will be counted as successful "green throws." Those which pass through the opening with their tip nearer the red side will be counted as successful "red throws." If one or both of the knives in any pair of knives fail to make it through the opening, the pair isn't counted when compiling the statistics.

---

[22]For the purposes of this illustration, it can be assumed that the twins ability to throw knives with matched pitch is due to the special way in which the knives are handed to the twins before each throw. This special method insures that the pitch of the knives is always random, but that both knives in any pair have exactly the same pitch. Also, in order for this demonstration to parallel as much as possible the typical correlation experiment, one should assume that, because the knives are thrown with great speed, both the pitch and roll of the knives remain essentially *fixed* while in flight. In other words, the knives remain fixed both as to their angle of pitch and their angle of roll from the moment they are released until the moment they reach their respective targets.

The center disk is once again set spinning and the blindfolded twins begin throwing successive pairs of knives with matched, albeit random, pitch -- just as they did in the previous portion of their act. Because they remain unable to see either target due to their blindfolds, the rate at which their successful throws are counted as green or red throws depends entirely on the random angle of roll of the knives when they reach their respective targets. In this part of the act, however, the target on the left remains stationary, fixed with its mark at the very top, and after each successive pair of knives is thrown, the target on the *right* side of the stage is rotated clockwise exactly one degree to a new position. This process is repeated until the target on the right has been rotated a full 180 degrees so that its mark – which had been at the top initially – is located at the target's bottom.

When the statistics are examined, no one is surprised. A graph showing how often the twins' throws were correlated with respect to whether they both were green or red throws – as the angle between the marks on the target openings was steadily increased – produces a straight line like the example shown below.

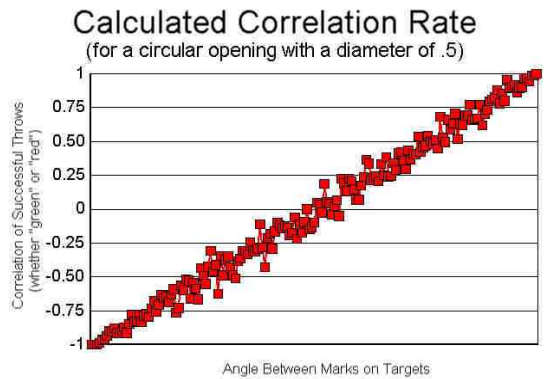

The straight line extends from a point showing complete *anti-correlation* when the targets were set to the same position (because the twins are still positioned head-to-toe *vis a vis* each other) – to a point showing complete *correlation* at the end of the demonstration when the targets were positioned opposite one another (i.e., with the left target still straight up at zero degrees and the right target positioned with its mark at 180 degrees, pointing straight down). The "zero correlation" value in the very middle of the graph – when the targets were positioned at 90 degrees perpendicular to one another – reflects the fact that, in that orientation, the throws were just as likely to be both green or both red as different from one another.[23]

Then, in Part Five of their act, the twins throw their knives in a way which – the announcer suggests – *can only be explained if the knives themselves are somehow **communicating** with each other while in flight*. The preparation for Part Five is nearly the same as before. The two small curtains are repositioned between the audience and each target, and the targets with the small circular openings are replaced with new targets. As before, the audience cannot see the new targets, but they are assured that each target has a small triangular mark at the top edge in the center, and the background of the targets remain colored just as before.

The blindfolded twins resume their positions strapped to the rotating disk, still positioned head-to-toe *vis a vis* each other. The rotating disk is then once again set rapidly spinning. Each target is then rotated to one of three separate random positions in preparation the first throw. Those three[24] random positions are identical for each twin: zero degrees (i.e., straight up), 22.5 degrees from vertical, and 67.5 degrees from vertical. Since the twins are facing opposite directions and are affixed to the same rotating disk, those target positions are measured clockwise from vertical for the twin on the left, and counterclockwise from vertical for the twin on the right so that the relative angles between the twin's orientation and the orientation of the marks at the top of the targets are the same for each twin *vis a vis* their own target.

---

[23]Note that the straight-line graph of the resulting correlation statistics would look the same even if the targets were randomly rotated to different positions during this part of the knife-throwing act, rather than being manually positioned for each throw so that the angle between the targets' respective axis of orientation increases gradually from a theta angle of zero to the final position where the theta angle is at its maximum.

The twins then perform just as they have before, throwing successive pairs of knives with perfectly matched, albeit random, roll and pitch. However, this time, between each successive throw, both targets are repositioned and set at random to one of the three optional positions. In order to inhibit the knives from being able to communicate with each other (see footnote 19 above), the targets are moved each time to their new random position very quickly *while the knives being thrown are actually in flight*.

A record of successful throws is kept, just as before, showing how often each twin throws a green throw, as opposed to a red one. Considered independently, the statistics for each twin appear completely random. In other words, there is absolutely no discernable pattern as to whether any successful throw will wind up being measured as green or red. Nevertheless, the overall statistical results of this demonstration astound every person in the audience at all familiar with the concept of quantum entanglement, because the results appear to prove that each pair of thrown knives must have been "entangled": **Whenever both targets happen to be set to the same optional position (for example, when *both* are set to the 22.5 degree angle position) the twins' successful throws are *always* perfectly anti-correlated with respect to whether the throws are green or red! In other words, when the targets are set to the same optional position, the two knives in any pair are *never* both green or both red.**

The demonstration is then repeated a large number of times using completely random angles for the positions of the targets anywhere throughout 360 degrees, and a graph showing the resulting overall statistics is prepared to depict the change in correlation rate as a function of the change from zero to 180 degrees in the theta angle between the small triangular marks on the edge of the target openings. That graph, reproduced below, traces out a perfect sine curve, *just like the results quantum theory predicts for correlation experiments using*

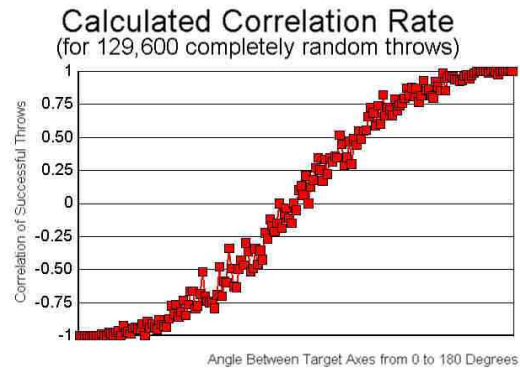

---

[24]The total number of separate random positions can be increased to any number without having any effect on the perception that the knives must somehow be communicating with each other during this portion of the twins' act. The only significance in increasing the total number of separate random positions to more than three is that a larger number of random positions tends to lessen proportionally the likelihood that the targets will happen to be set to the *same* position for a particular throw, thereby increasing the time it takes to gather useful correlation data.

*spin-½ particles*,[25] despite the fact that the knives in this demonstration can be regarded as completely unconnected independent physical entities, none of which should be able to communicate with any other knife or other object in any way.

What can explain – everyone asks – how the knives can "know" whether they should be oriented to count as a green or red throw in order to properly correlate the number of green and red throws to produce the statistics predicted by quantum theory? Neither blindfolded twin can know which position either target disk will be set to when the knives are thrown, since the targets aren't set until *after* the knives are thrown, yet still the green/red results are perfectly anti-correlated whenever the targets wind up being set to the same position. The only possible explanation, one member of the audience insists, is that each knife must somehow be *preprogrammed* appropriately to insure that it will be green or red as needed for each possible optional target position. Furthermore, at least one of the knives must then find a way to inform its mate – while in flight – how the first knife's target is oriented (i.e., to which of the optional positions that target is set) before the second knife reaches *its* target. Then, having received that information, the second knife must somehow adjust its roll and pitch accordingly to insure that the expected green/red statistics are generated. If there are three optional settings, simple logic – he insists – requires that those programming instructions consist of no less than three bits of information:[26]

(1) Whether it will have to be oriented as a green throw if the other target is set to the first optional position; (2) Whether it will have to be oriented as a green throw if the other target is set to the second optional position; and (3) Whether it will have to be oriented as a green throw if the other target is set to the third optional position.

After much consternation and debate, however, the announcer pulls back the curtains and reveals the secret behind the twins' amazing performance: *Their performance was due, once again, to the unusual **shape** of the target openings used in this part of their act.* This time, instead of using targets with narrow slits, or ones with circular openings, the targets had a figure-eight shaped opening similar to the drawing depicted below.[27]

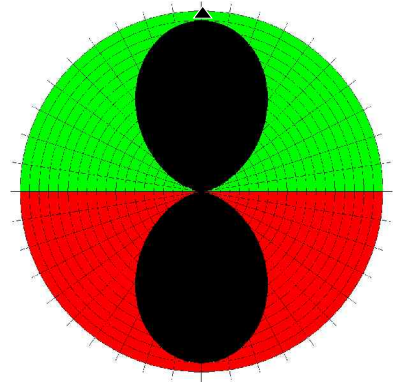

---

[25]The non-linear function quantum theory predicts for spin-½ particles in this type of correlation experiment is -cos θ. For spin-1 particles, quantum theory predicts a similar non-linear correlation function of $\cos^2 θ$.

[26]It is worth noting that the number of optional target positions is essentially infinite if the targets are rotated at random to *any* position within their full 360 degrees of rotation, making the task of preprogramming in such a scenario exceedingly challenging.

[27]The figure-eight shape in the center of the target is a polar graph of the equation "½ [1+cos(2θ)]", where *theta* (θ) is the angle in degrees measured from the vertical. The drawing shows how the targets were oriented in space *vis a vis* the twins at the start of this portion of their act. The small triangular mark on the top edge of each target was positioned at the center of the figure-eight shape at zero degrees, with the maximum clearance through each opening along that top-to-bottom axis equal to the overall length of each knife.

To understand how the use of this figure-eight shaped target generated such curious correlation statistics, consider what happens when "entangled" knives are thrown at targets with this shape of opening: If the two targets are aligned with each other, *both* knives in any pair will either pass successfully through the opening or fail to pass through the opening, since both of the knives in any "entangled" pair have a *matching* angle of pitch and an angle of roll which is essentially the same (their angle of roll is actually 180 degrees apart because the twins are positioned head-to-toe *vis a vis* each other, but the figure-eight shaped target looks exactly the same when turned upside down).

Because the angle of roll for "entangled" pairs of knives is always 180 degrees apart, pairs which successfully pass through their respective targets will be perfectly *anti-correlated* with respect to whether they are counted as green or red throws (i.e., one will always be green, whereas the other will always be red) – regardless of their angles of pitch and roll – *as long as the targets remain aligned with one another*. If the targets are repositioned so that the small marks at the top of each target are 180 degrees *opposite* one another, successful throws will be perfectly *correlated* with respect to whether they are counted as green or red throws, since in that orientation each target will be positioned identically from the viewpoint of the twins.

If the angle between the marks at the top of the targets is set to some intermediary angle so the longitudinal axes of the target openings are no longer aligned, the matched pitch and opposite angle of roll of "entangled" pairs of knives will insure that only *some* of the knives thrown at the targets will pass through their respective target openings.

Consider, for example, what happens if the twins resume their starting position and the left target remains positioned with its small mark located at zero degrees as shown in the drawing on the left at the top of the next page, but the target on the right is positioned so that its mark is at 60 degrees counterclockwise from vertical. *From the viewpoint of the twins* (recall that the twin on the right is upside down at this point), the targets would look like the drawings shown at the top of the following page, with the vertical white line through the middle indicating the available *clearance* through the openings with respect to each twin.

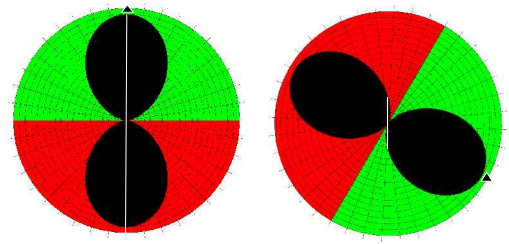

Left twin's view of his target     Right twin's view of his target

Because all of the knives in the foregoing illustration were thrown in a way which guaranteed that their angle of their *pitch* would at all times be totally random (albeit matched), pairs of knives thrown with a perfectly vertical angle of roll – matching the orientation of the white line in the drawings shown above – both passed through the figure-eight shaped openings and were counted only if the angle of pitch was sufficiently horizontal (like a dart sticking in a dartboard) to allow the knife thrown at the target on the right to fit through the available clearance (shown by the shorter white line in the center of that target opening). Where the theta angle between the small marks at the top of the target openings is equal to 60 degrees, only 25% of "entangled" pairs of knives thrown from the aforementioned starting position with random pitch at targets in that orientation *vis a vis* one another will pass through both openings.[28]

Additionally, with the targets positioned as shown above, a significant portion of the time – depending on the random angle of roll (which can be anywhere throughout 360 degrees) – *both* knives which happen to pass through their respective targets will be counted as green (or red). That percentage of correlated green/red throws increases as the theta angle between the marks at the top of the targets increases from zero to 180 degrees. At the point where the marks on the top of the targets are positioned 180 degrees opposite one another, fully 100% of successfully thrown knives will be identified as the same color throw (i.e., their color designation will be completely correlated), since at that point, the targets are oriented identically *vis a vis* the twins.

Because the pitch of any pair of knives, although fixed while in flight, is completely random, the *overall* probability that knives will be thrown with a pitch sufficient to allow the pair of knives to pass through their respective openings is determined strictly by the figure-eight shaped openings in the targets. Due to that shape, the overall rate of success in passing through the figure-eight shaped openings will always be close to 50%.

As expected, the statistical results of this sort of correlation experiment range from complete *anti*-correlation – when the targets are aligned with each other – to complete *correlation* of results when the targets are oriented 180 degrees opposite one another. When the targets are oriented *perpendicular* to one another, there is no correlation whatsoever (i.e., the odds are 50/50 that any "entangled" pair will be measured as both being the same color throw, as opposed to different colors). A graph of the statistical results produced in this part of the knife-throwing experiment -- calculated using the spreadsheet found in Appendix A -- looks like the drawing shown below.

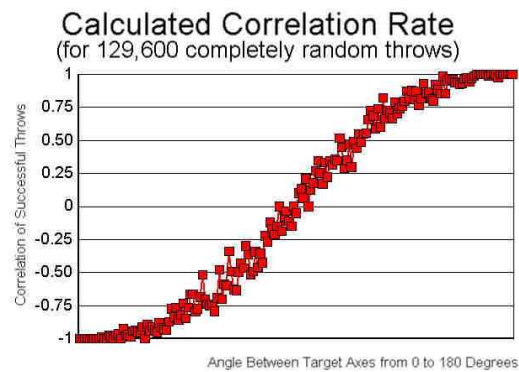

The spreadsheet in Appendix A calculates the rate of correlation of green/red results using completely random pitch and roll variables for the knives. The orientations of the targets also range throughout a complete 360 degrees insuring essential randomness. A calculation is made for each separate pair of thrown knives to determine whether they both would have passed through their respective targets, and for those occasions where both throws would have been successful, a separate calculation is then made to determine whether each knife would have been measured as a green throw or a red throw.[29] That determination depends on their random degree of roll *vis a vis* their respective targets. As note above, the overall statistical results produce the kind of correlation curve quantum theory predicts for the typical EPR-type correlation experiment.

---

[28] The percentage of randomly thrown knives which will pass through this figure-eight shaped opening at a given angle can be determined by comparing the available clearance at that angle to the length of the knives (which is equivalent to the maximum clearance). The mathematical formula for the shape of this target opening, "½ [1+cos(2θ)]", can be used to calculate the distance across the opening at the specified angle -- but note that the θ angle symbol in the formula refers to the angle measured **from the x-axis** (i.e., from the axis separating green from red). For example, using the orientation of the target on the right shown above, the θ angle to be inserted into the formula is *30* degrees, representing the angle between the x-axis and the white line shown on the target on the right. The calculated length of the white line shown in that target on the right (the available clearance through that opening) works out to be 25. Since the white line on the left is 100% of the maximum length, the shorter white line is 25% of the length of the knives. Consequently, there is a 25% probability that pairs of "entangled" knives thrown with random pitch at that specified angle of roll at targets positioned as shown above will pass through both openings.

This explains, then, how the twins were able to make it appear that their knives were communicating with each other in flight: The relative probability of passing through the target openings and producing, in turn, the sinusoidal correlation statistics predicted by quantum theory was determined by the shape of the target openings themselves, not by any mystical collaboration between the knives or other "spooky action at a distance." As with the menus involved in the earlier illustration about the two brothers who ate in different restaurants, the shape of the target openings in the knife-throwing illustration played a direct, active role in *determining* – not just recording or reporting – the overall statistical results.[30]

Of course, the shape of the target openings did not play the *only* role in determining the resulting correlation statistics. The experimental results in both scenarios (i.e., whether certain food items were ordered by the brothers, or in the second illustration involving the knife-throwers, whether certain successful throws were counted as "green" or "red" throws) were influenced in part by the random

---

[29] It is worth recalling at this point that although the original "thought experiment" proposed by Einstein in his famous EPR paper envisioned experiments which measured the location and momentum of spin-½ particles, the correlation experiments which have been conducted in the last several decades typically involve measurements of the polarization of photons, which are spin-1 particles. To produce statistical results in the knife-throwing illustration identical to the results obtained in experiments using photon polarization, one can simply have the twins repositioned back-to-back instead of head-to-toe *vis a vis* each other (since pairs of photons which are "entangled" using the process discussed above invariably have the *same* polarization rather than the *opposite* polarization) and then perform the knife-throwing demonstration using a four-petaled rhodon target shape which looks like the drawing shown below.

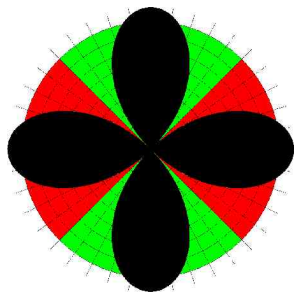

The opening in this target is simply a polar graph of the mathematical formula: "$\cos(2\theta)$". The colored sections are split into *four* equal sized areas instead of just *two* to more closely approximate the way photon-based correlation experiments are performed. In photon-based correlation experiments, the statistics in question involve whether or not photons are polarized in a vertical – as opposed to horizontal – orientation. A graph of the statistical results produces the same sinusoidal curve, except that it appears flipped on its horizontal axis. Consequently, if targets like those shown above are aligned with each other in a knife-throwing experiment, the results are perfectly correlated. When the targets are *perpendicular* to one another, the results are perfectly *anti-correlated*.

[30] With respect to the illustration involving restaurants, the overall likelihood of correlated results when the brothers do *not* happen to eat at the same restaurant depends on the extent to which identical food items are offered on different menus, and precisely where those items appear on the menus. The fewer choices there are, and the more similar the different menus are to each other with respect to the positioning of identical food items, the higher the correlation will be between the choices made by the brothers when eating at *different* restaurants on the same day.

data unique to the "entangled" objects themselves (i.e., the random numbers chosen by the brothers, or the random orientation in three-dimensional space of the knives). Nevertheless, neither the menus in the first illustration nor the shape of the openings in the targets in the second acted merely as a passive means of receiving and reporting results which had already been entirely "preprogrammed." If the content of the menus had been modified while the brothers were traveling to their respective restaurants, or if the shape of the target openings in the knife-throwing illustration had been modified while the knives were traveling through the air to their respective targets, the resulting correlation statistics very likely would have been different, even though the things that were supposedly "entangled" (i.e., the brothers eating lunch in the first illustration, or the knives in the second illustration) were not changed in any way!

This shows, then, why one cannot simply assume that the experimental apparatus in correlation experiments plays a completely passive role in determining the statistical results. The Passive Apparatus Assumption is obviously *not* a valid assumption in the knife-throwing illustration -- and Bell's Inequality cannot be properly utilized to analyze the results of such a correlation experiment unless the Passive Apparatus Assumption is *true*.

Again, Bell's Inequality is premised on the assumption that the hidden variables, if there are any, are attributes intrinsic to the "entangled" objects themselves, as opposed to being attributes of the apparatus used to measure the "entangled" objects. If there are "hidden variables" in the experimental apparatus affecting the statistical outcome of the experiment, Bell's Inequality is of absolutely no use in excluding the possibility that there are *additional* variables hidden within in the "entangled" objects themselves which also affect those statistics.

The foregoing discussion also shows why the related Minimum Information Assumption – the assumption that the phenomenon of quantum entanglement absolutely requires entangled objects to be preprogrammed at the time they are created with separate bits of information numbering at least as many as the optional target settings which the entangled objects might encounter in the particular experiment in question – must also be considered false whenever the Passive Apparatus Assumption is false.

Of course, if the figure-eight shape of the target openings in the foregoing knife-throwing illustration were an unusual shape not likely to be found in nature, or if the figure-eight shaped curve used for the target openings had no relationship to the actual physical processes taking place when light passes through a polarizing filter (or the process involved in determining particle spin using Stern-Gerlach magnets in a correlation experiment using spin-½ particles), one might not be inclined to consider the knife-throwing illustration reasonably analogous to the typical correlation experiment. However, as discussed below, the figure-eight shaped curve used for target openings in the knife-throwing illustration bears a strong resemblance to the shape – at a subatomic level – of the "targets" used in the typical correlation experiment.

b. The Nature of Photon "Targets" Used in Correlation Experiments

Photon-based correlation experiments all depend on determining the polarization of pairs of "entangled" photons in one way or another. Polarization of photons is often determined using some sort of polarizing filter or similar optical device. A generic polarizing filter is manufactured by stretching a transparent sheet of long, string-like polymer molecules, such as poly(vinyl alcohol), and then immersing the sheet in an iodine solution so the iodine molecules are absorbed into the sheet.

The iodine atoms, which are electrophilic due to their large, polarizable outer electron clouds, accumulate as polyiodine complexes, forming long linear chains which line up parallel to the strands of polymer. Those chains then act like numerous incredibly thin *wires*, allowing electrons to move along those wires parallel to the axis along which the polyiodine complexes have aligned themselves.

Considered individually, iodine atoms have a total of 53 electrons which are arranged in a series of orbitals surrounding the nucleus, all of which are completely filled with electrons, except for a single one of the three outer $5p$-orbitals. The one unfilled outer $5p$-orbital contains only one of the two electrons which can fill that orbital.[31] The unfilled orbital is what makes the atom polarizable, causing the atom to align in the same three-dimensional orientation with all of its neighboring iodine atoms on the axis along which the polymer molecules have been stretched.

In a photon-based correlation experiment using this type of polarizing filters, the filters are struck by approaching "entangled" photons, allowing only photons whose axis of linear polarization is within a certain range to pass through the filter. Linearly polarized photons of light whose undulating electric field is *parallel* to the iodine "wires" cause movement of electrons encountered in the unfilled outer $5p$-orbital in one iodine atom to the same orbital in the next iodine atom in the chain, creating an electrical current. The resistance to current flow in those "wires" causes the moving electrons to lose energy in the form of heat. The photons of light which are absorbed by the iodine in that process are thereby prevented from passing through the polarizing filter.

Photons whose electric field is more or less *perpendicular* to the axis along which the "wires" are aligned tend *not* to be absorbed, since the electrons in the iodine atoms in that orientation – which occupy the fully filled orbitals – are relatively immobile. In other words, photons in the typical correlation experiment which have an axis of polarization more or less *perpendicular* to the axis of polarization of their respective polarizing filter pass through the filter and on to the correlation counter. The correlation counter then tallies up the number of times that pairs of "entangled" photons arrive at the same instant at the correlation counter, thereby generating statistics showing the number of "entangled" pairs which have correlated polarization for the particular theta angle at which the polarizing filters are set with respect to each other.

---

**31**The three separate $5p$-orbitals in individual iodine atoms are oriented perpendicular to each other as shown in the drawing shown at the bottom of this column on the left, although the actual three-dimensional shape of those orbitals is significantly more complex than the simple figure-eight representations often depicted. The actual shape of one of those orbitals looks more like the example shown below on the right, with the two orbital lobes shown in different colors. Also, unlike the drawing shown below on the left, each orbital is obviously centered about the same point in space:

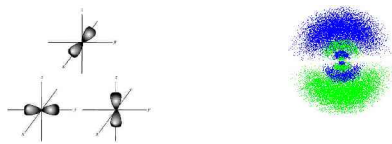

From the viewpoint of "entangled" photons in the typical correlation experiment, the iodine atoms found in polarizing filters must look very much like the figure-eight shaped target openings confronting the knives in the knife-throwing illustration, except that in the knife-throwing illustration, there is only a single target opening on each side of the "apparatus," whereas in the typical photon-based correlation experiment, there are huge numbers of individual iodine atoms – each functioning as a separate "target." That difference, however, obviously wouldn't significantly affect the overall correlation statistics, since the millions of individual iodine atoms are all *aligned* with each other, so that they look similar to the vertical strings of small figures shown in the drawing below.

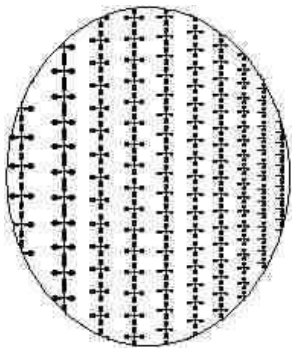

Put another way, one would not expect the overall statistical results to be any different in the knife-throwing illustration if each pair of "entangled" knives were thrown at *separate* target openings, as long as each of those many separate openings were aligned with one another in a similar manner.

Although, considered individually, the complex three-dimensional shape of a 5*p*-orbital in an iodine atom is obviously not identical to the figure-eight shaped target opening in the knife-throwing illustration, the shape of the target openings and the shape of the 5*p*-orbitals in iodine atoms found in polarizing filters both can be viewed as identifying the "polarity" of the objects they encounter. The likelihood of a photon encountering an electron in the outer 5*p*-orbital of an iodine atom in a polarizing filter and as a result being prevented from passing on to the correlation counter depends – at least to a significant degree – upon the photon's axis of polarization (i.e., its angle of "roll" in three-dimensional space *vis a vis* the filter). Similarly, the likelihood of knives being blocked by the orbital-like shape of the target openings in the knife-throwing experiment also depends to a significant degree upon the knives' angle of roll. Since that likelihood, in both cases, conforms to Malus' Law, and since the mathematical formula upon which that law is based matches precisely the curve of the target openings, the knife-throwing illustration seems reasonably analogous to the typical EPR-type correlation experiment.[32]

Although we will likely never be able to "see" the actual physical structure of photons or other subatomic particles – or know for sure what physical processes cause them to interact the way they do with polarizing filters or other kinds of experimental apparatus – one thing is certain: the knife-throwing illustration clearly shows that it *is* possible to produce statistical correlation results in a purely classical way so that they match *exactly* the predictions made by quantum theory for the expected results of a typical EPR-type correlation experiment. The key point is that such a feat is possible only if the predicted probability of correlated results is *determined by* – not just measured by – the experimental apparatus. The hidden variable, in short, has been hiding within the apparatus, not within the particles being measured.[33]

---

[32] As noted above, correlation experiments can theoretically also be conducted using Stern-Gerlach magnets to detect the *spin*, as opposed to the *polarization*, of spin-½ particles. A similar quantum mechanical Malus' Law holds in those sorts of correlation experiments – the only difference being that the formula differs by a factor of ½. In other words, instead of the formula "$\cos^2 \theta$", Malus Law as it pertains to correlation experiments involving the spin of spin-½ particles uses the formula "$\cos^2 \theta/2$" to calculate the change in probability of encountering a specified particle spin as the orientation of the magnets are rotated.

Because the Passive Apparatus Assumption is demonstrably false as it relates to the knife-throwing illustration, it seems logical to assume that that assumption may well be equally false as it relates to the typical EPR-type correlation experiment. If the experimental apparatus in the typical photon-based correlation experiment is *not* completely passive, as the researchers who have conducted those experiments – and who have used Bell's Inequality to analyze the results of those experiments – have all uniformly assumed, rejecting the Passive Apparatus Assumption when analyzing the statistical results of correlation experiments can lead to a new, fundamentally *deterministic* view of the nature of physical reality.

It is certainly not difficult to see why conventional wisdom has always assumed that a photon's likelihood of passing through a polarizing filter is based solely on its "roll variable" (i.e., its orientation in three-dimensional space perpendicular to its direction of travel and perpendicular to its postulated "pitch variable"), and that that variable is *probabilistic*, rather than *deterministic*. Bell's Inequality firmly establishes that no variable *intrinsic to the photons themselves*, including any so-called "pitch variable", can be responsible for producing the bizarre statistics generated in the typical correlation experiment.

However, if one accepts the fact that the experimental apparatus can play more than a purely passive role in determining which photons pass through polarizing filters, and if one recognizes, as Bell himself seemed to at one point,[34] that the Passive Apparatus Assumption is fundamentally inconsistent with quantum theory's teachings about the nature of our universe, it becomes easy to see how the phenomenon of quantum entanglement can be understood in a way which is fully consistent with relatively theory and the Locality Principle.[35]

---

[33] In the last couple of decades, a number of writers have questioned the generally accepted view that the statistical results of EPR-type correlation experiments prove Einstein must have been wrong in believing in the Locality Principle. Those efforts, however, have mostly focused on questioning whether the experimental apparatus can adequately detect and tally all of the correlated pairs of "entangled" photons generated in the experiment. None of those writers appear to have considered the possibility that the experimental apparatus itself plays a central role in producing the sinusoidal statistical results predicted by quantum theory.

[34] Ironically, Bell seems to have anticipated the ideas discussed herein when he wrote the following about "measuring" things in a quantum mechanical world: "The word very strongly suggests the ascertaining of some preexisting property of some thing, any instrument involved playing a purely passive role. Quantum experiments are just not like that, as we learned especially from Bohr. The results have to be regarded as the joint product of 'system' and 'apparatus,' the complete experimental set-up. But the misuse of the word 'measurement' makes it easy to forget this and then to expect that the 'results of measurements' should obey some simple logic in which the apparatus is not mentioned."

[35] Appendix B contains a spreadsheet with formulas allowing one to calculate the extent to which the results of a correlation experiment like the above-referenced knife-throwing experiment violate the CHSH version of Bell's Inequality. As with the other spreadsheet provided in Appendix A, the roll and pitch of each and every

### 3. Conclusion - The Road Ahead

The phenomenon of quantum entanglement has always been perplexing, because conventional analysis of correlation experiments – an analysis which relies upon a flawed application of Bell's Inequality – leads to the counter-intuitive conclusion that measurement of a photon's polarization in one location can have an instantaneous effect on the polarization of an "entangled" photon in some other distant location, despite what Einstein's Special Theory of Relativity says about the impossibility of instantaneous action at a distance. The mistaken view that Einstein has been proven incorrect by the results of EPR-type correlation experiments can be shown to be the product of an unjustified assumption regarding the role which the experimental apparatus plays in such experiments. If one recognizes and then sets aside the erroneous assumption that the experimental apparatus in such experiments is necessarily purely passive, the phenomenon of quantum entanglement can be explained, as it has been above, in a purely classical way which does not require instantaneous action at a distance.

Not surprisingly, this new, untested and relatively heretical explanation of quantum entanglement gives rise to a number of new concerns, such as how the "pitch" of a photon can be measured or manipulated, and whether there remains any hope that the phenomenon of quantum entanglement can still form the basis for practical applications in the fields of quantum encryption and quantum teleportation. Much work also obviously needs to be done to determine the extent to which the ideas expressed herein may explain, if at all, other phenomena, such as the "entanglement" effects perceived in experiments using other types of subatomic particles and other testing methodologies. Since, however, all correlation experiments make use of apparatus made of atoms of one sort or another, and since the shape of the atomic orbitals in the atoms within those portions of the experimental apparatus used to measure spin, polarization or other similar variable – as well as those atoms' spacial orientation with respect to one another – will likely be analogous to some substantial extent to those found in polarizing filters, it does not seem unreasonable to anticipate that continued research into this fascinating area of physics – coupled with the recognition that a correlation experiment's apparatus can have a direct, causal effect on the results of the experiment – will eventually advance our understanding of quantum phenomena to the point that the scientific community will be able to say with great confidence that quantum theory was clearly "incomplete" at the point in time when Einstein published his EPR paper -- at least insofar as it ever accepted the reality of instantaneous action at a distance in our universe. When and if that point is reached, it will be a tremendous tribute to Einstein's remarkable intuition and his incomparable genius.

---

"entangled" object are completely random, albeit matched with respect to entangled pairs. The statistical results produced by both of those two spreadsheets reflect precisely the values predicted by quantum theory.

# APPENDIX

**Note regarding Appendix A and Appendix B to this paper:** The complete spreadsheets referenced above in the body of this paper (which are several megabytes in size), as well as a number of related mathematical calculations, together with a complete copy of this paper containing high quality graphics, can be downloaded from:

http://www.intelligent-tech.com/docs-jlf

**References:**                                             [*] Author's email address: resipsaloquitur@msn.com

1. A. Einstein, *Zur Electrodynamik bewegter Körper*, Annalen der Physik, 17:891 (1905).

2. A. Einstein, N. Rosen and B. Podolsky, *Can Quantum-Mechanical description of physical reality be considered complete?*, Phys. Rev. 47, 777 (1935).

3. J. S. Bell, Physics 1, 195 (1965), reprinted in J. S. Bell, *Speakable and Unspeakable in Quantum Mechanics*, (Cambridge University Press, Cambridge, 1987).

4. J. F. Clauser, M. A. Horne, A. Shimony and R. A. Holt, *Proposed experiment to test local hidden-variable theories*, Phys. Rev. Lett. 23, p. 880 (1969).

5. A. Aspect, P. Grangier and G. Roger, *Experimental Realization of Einstein-Podolsky-Rosen-Bohm Gedankenexperiment: A New Violation of Bell's Inequalities*, Phys. Rev. Lett. 49, p. 91 (1982).